\documentclass[sigconf]{acmart}
\AtBeginDocument{%
  }

\usepackage{multirow}
\usepackage{multicol}
\usepackage{makecell}
\usepackage{balance}
\copyrightyear{2025}
\acmYear{2025}
\setcopyright{acmlicensed}
\acmConference[SIGIR '25] {Proceedings of the 48th International ACM SIGIR Conference on Research and Development in Information Retrieval}{ July 13--18, 2025}{Padua, Italy.}
\acmBooktitle{Proceedings of the 48th International ACM SIGIR Conference on Research and Development in Information Retrieval (SIGIR '25), July 13--18, 2025, Padua, Italy}
\acmDOI{10.1145/3726302.3730253}
\acmISBN{979-8-4007-1592-1/2025/07}

\begin{document}

\title{Template-Based Financial Report Generation in Agentic and Decomposed Information Retrieval}


\author{Yong-En Tian}
\authornote{Both authors contributed equally to this research.}
\affiliation{
    \institution{National Yang Ming Chiao Tung University}
    \department{Department of Computer Science}
    \city{Hsinchu}
    \country{Taiwan}
}
\email{bryanttian.cs12@nycu.edu.tw}

\author{Yu-Chien Tang}
\authornotemark[1]
\affiliation{
    \institution{National Yang Ming Chiao Tung University}
    \department{Department of Computer Science}
    \city{Hsinchu}
    \country{Taiwan}
}
\email{tommytyc.cs10@nycu.edu.tw}

\author{Kuang-Da Wang}
\affiliation{
    \institution{National Yang Ming Chiao Tung University}
    \department{Department of Computer Science}
    \city{Hsinchu}
    \country{Taiwan}
}
\email{gdwang.cs10@nycu.edu.tw}

\author{An-Zi Yen}
\affiliation{
    \institution{National Yang Ming Chiao Tung University}
    \department{Department of Computer Science}
    \city{Hsinchu}
    \country{Taiwan}
}
\email{azyen@cs.nycu.edu.tw}

\author{Wen-Chih Peng}
\affiliation{
    \institution{National Yang Ming Chiao Tung University}
    \department{Department of Computer Science}
    \city{Hsinchu}
    \country{Taiwan}
}
\email{wcpeng@cs.nycu.edu.tw}

\renewcommand{\shortauthors}{Yong-En Tian, Yu-Chien Tang, Kuang-Da Wang, An-Zi Yen, \& Wen-Chih Peng}

\begin{abstract}
Tailoring structured financial reports from companies' earnings releases is crucial for understanding financial performance and has been widely adopted in real-world analytics.
However, existing summarization methods often generate broad, high-level summaries, which may lack the precision and detail required for financial reports that typically focus on specific, structured sections.
While Large Language Models (LLMs) hold promise, generating reports adhering to predefined multi-section templates remains challenging.
This paper investigates two LLM-based approaches popular in industry for generating templated financial reports: an agentic information retrieval (IR) framework and a decomposed IR approach, namely AgenticIR and DecomposedIR. 
The AgenticIR utilizes collaborative agents prompted with the full template.
In contrast, the DecomposedIR approach applies a prompt chaining workflow to break down the template and reframe each section as a query answered by the LLM using the earnings release.
To quantitatively assess the generated reports, we evaluated both methods in two scenarios: one using a financial dataset without direct human references, and another with a weather-domain dataset featuring expert-written reports.
Experimental results show that while AgenticIR may excel in orchestrating tasks and generating concise reports through agent collaboration, DecomposedIR statistically significantly outperforms AgenticIR approach in providing broader and more detailed coverage in both scenarios, offering reflection on the utilization of the agentic framework in real-world applications.
\end{abstract}




\maketitle
\vspace{-5pt}
\section{Introduction}
Crafting a comprehensive financial report is an essential task for analysts to perform inter-company financial performance evaluation, and the resulting assessment report can be employed in a wide range of applications, e.g., stock movement prediction \citep{ECHO, stock_price_movements}, financial risk prediction \citep{hypergraph, DialogueGAT}, and financial sentiment analysis \citep{Du_financial_sentiment}.
With the fruitful information disclosed in earnings call statements, the financial report can effectively facilitate not only the analysts' and investors' decision-making \citep{keith-stent-2019-modeling}, but also competitive companies' understanding of the market dynamics \citep{adom2016competitor}.
Despite its broader application, tailoring a well-organized financial report is a challenging and time-consuming task for analysts, as it demands the careful selection of financial evidence from lengthy earnings call documents \citep{finance_bench}.
A common approach is to frame it as an earnings call transcript summarization task \cite{ECTSum, insturction_guided_bullet}.
Several methods employing AI models to tackle this problem have been proposed \citep{ECTSum, insturction_guided_bullet, extractive_summarization}; these methods mainly seek to extract relevant information from the earnings call transcript and organize it into a compelling report.

\begin{figure}[t]
    \centering
    \includegraphics[width=0.75\linewidth]{
    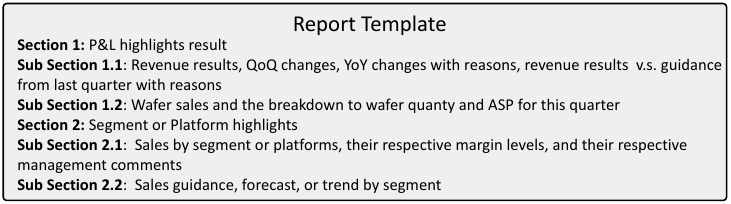}
    \vspace{-5pt}
    \caption{An example of the financial report template.}
    \vspace{-10pt}
    \label{fig:template}
\end{figure}

Recently, LLMs have become a promising approach to solve the summarization task in an agentic manner, where the LLMs are instructed to follow a pre-defined procedure to solve multi-step tasks \citep{Dohan_langcascade, wu2024stateflow}.
For instance, \citet{from_fact_to_insights} proposed a multi-agent framework to collaboratively generate insightful reports from earnings call transcripts.
In addition to agent-based approaches, some studies have begun to explore specific strategies to improve summarization performance. 
\citet{multi-agent_rag} introduced a framework whereby multiple LLMs autonomously retrieve knowledge and generate context-aware responses to address complex, dynamic decision-support tasks. 
Alternatively, \citet{rag_based_summary} focused on decomposing the summarization task into a series of prompt-based subquestions, integrating RAG \citep{rag} to improve modularity and control over content. 
Despite the promise of these approaches, there has been limited exploration of their comparative effectiveness in terms of generating financial reports, particularly when aligning with expert-defined templates as shown in Figure \ref{fig:template} or decomposing templates into subqueries for more detailed insights.

This study aims to explore the differences between two approaches for template-based financial report generation: Agentic Information Retrieval (AgenticIR) and Decomposed Prompting \citep{prompt_chaining} Information Retrieval (DecomposedIR). 
In the AgenticIR approach, the LLM acts autonomously, directly utilizing a report template to retrieve and generate a financial report, which enables end-to-end report generation without explicit template-to-query decomposition.
In contrast, the DecomposedIR approach follows a prompt chaining workflow to break down the report template into multiple subqueries, each guiding the retrieval and generation process step by step. 
We hypothesized that while AgenticIR may excel in its flexibility, the DecomposedIR approach could yield more detailed and contextually rich financial reports by breaking down the task into manageable sub-modules.
\begin{figure*}[t]
    \centering
    \includegraphics[width=0.75\linewidth]{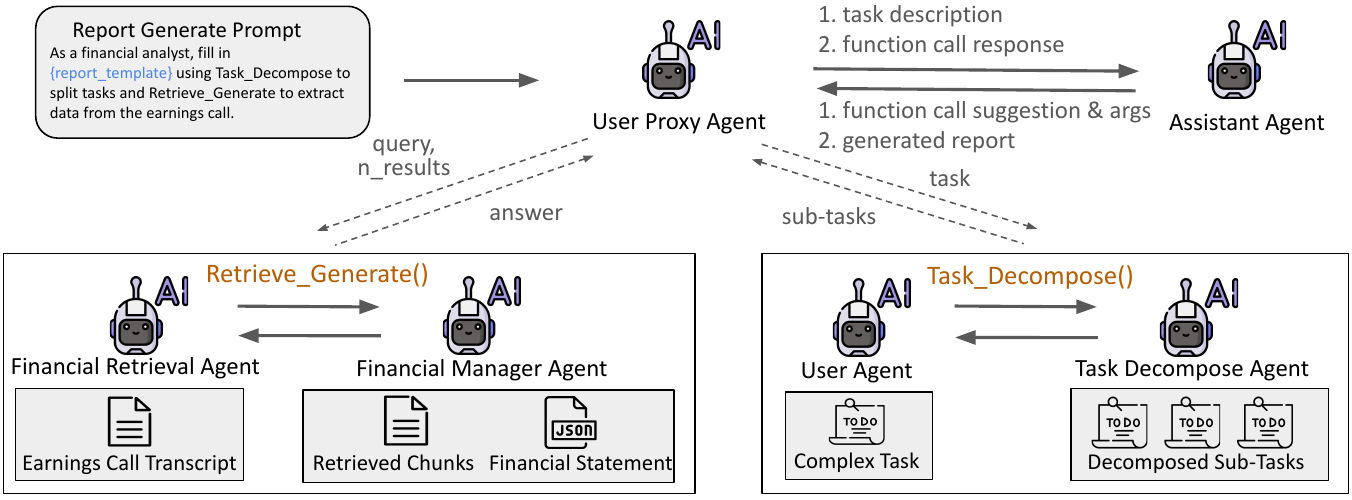}
    \vspace{-5pt}
    \caption{An illustration of the AgenticIR framework.}
    \vspace{-5pt}
    \label{fig:autogen}
\end{figure*}

To the best of our knowledge, this is the first systematic comparison of the quality of template-based financial reports generated by the AgenticIR framework and the DecomposedIR  approach.
Our contributions are threefold: 
(1) We introduce the task of template-based financial report generation and leverage the capabilities of LLMs to address it.
We then conduct a detailed comparative analysis of two predominant approaches in industries, AgenticIR and DecomposedIR, to elucidate their nuanced differences.
(2) Through extensive experiments, we demonstrate that despite the flexibility of the agentic framework, decomposed prompting consistently produce more detailed reports in the financial domain dataset across 4 key characteristics.
Moreover, our experiments on a weather domain dataset yield similar findings, underscoring the potential for cross-domain generalizability. 
(3) Using our findings, we compile insights to aid practitioners into the deployment of agentic framework in industrial settings.

\vspace{-5pt}
\section{Related work}



With increasing research and applications of LLMs as autonomous agents \cite{xi_llm_agent_survey, wang_llm_agent_survey}, numerous frameworks have been proposed to enhance their collaborative and problem-solving capabilities. 
CAMEL \cite{camel} introduced a role-playing framework that enables autonomous cooperation between communicative agents.
AutoGen \cite{autogen} provides a multi-agent conversation framework that enables agents to collaborate through LLMs, human inputs, and tools. 
LLMCompiler \cite{llm_compiler} enhances agent capabilities by optimizing the execution of parallel functions. 
\citet{from_fact_to_insights} is closely related to our work, as it explored the impact of the multi-agent framework on the generation of earnings call transcript summaries. 
However, their focus was primarily on the textual content generated by the agents, while our research investigated the effectiveness and difference of AgenticIR compared to RAG with decomposed prompting in template-based report generation.
\vspace{-5pt}
\section{Method}
Financial analysts need to analyze multiple companies' earnings call transcripts on a quarterly basis to gain insights into each company's operating performance, and to make informed decisions for future planning. 
However, traditional summarization methods often generate overly broad results, lacking the depth and detail required to address the specific topics financial analysts are most interested in. 
To address this issue, we reframed the summarization task as a template-based question-answering task, with the aim of concisely extracting information aligned with these topics.

\vspace{-5pt}
\subsection{Data Collection and Preprocessing}
As a pilot exploration, we selected five leading semiconductor manufacturing companies---TSMC, Intel, Samsung, GlobalFoundries, and UMC---and collected their earnings call documents through the Discounting Cash Flows API.\footnote{\url{https://discountingcashflows.com/}}
Specifically, we gathered the available earnings releases from 2021 to 2024, resulting in a dataset of 74 financial documents.
These documents encompassed the earnings call transcripts and the financial statements, which contain key financial data such as balance sheets, cash flow statements, and income statements. 
We preprocessed both types of document before applying the AgenticIR and DecomposedIR approaches. 
First, the transcripts were divided into chunks of 1,000 characters, with a 200-character overlap between consecutive chunks to preserve contextual coherence while ensuring manageable sizes for processing. 
These chunks were then used as input for retrieval in both approaches.
The financial statements were structured in JSON format and contributed to complementing the numerical information retrieved from the transcripts.
Furthermore, we collected the financial report template from a financial analyst and applied it throughout our experimental setup. 
The results are publicly available.\footnote{\url{https://github.com/bryant-nn/Template-Based-Financial-Report-Generation}}

\begin{figure}[t]
    \centering
    \includegraphics[width=0.75\linewidth]{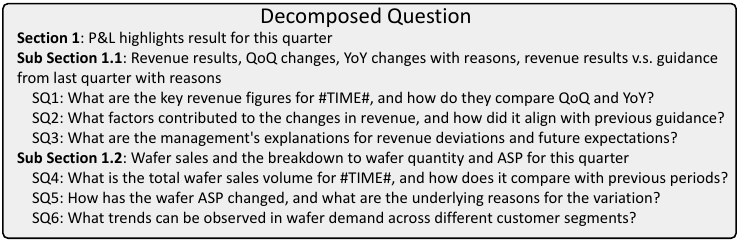}
    \vspace{-5pt}
    \caption{An example of the derived decomposed questions.}
    \vspace{-10pt}
    \label{fig:decomposed}
\end{figure}
\vspace{-5pt}
\subsection{AgenticIR}
We designed a multi-agent framework that utilizes two core functions, task\_decompose and retrieve\_generate, to collaboratively generate structured financial reports. 
These functions enable the agents to break down the task into manageable subtasks and retrieve relevant information for report generation.
We implemented this framework using AutoGen \citep{wu2024autogen}.
As illustrated in Figure \ref{fig:autogen}, the framework comprises six agents and two core functions. 
The six agents include a user proxy agent, assistant agent, financial retrieval agent, financial manager agent, user agent, and task decomposition agent. 
The two core functions represent interaction protocols between specific agents, where their outputs serve as the results of their interactions.

\noindent \textbf{Workflow overview.}
The process begins when the user proxy agent receives a report prompt and a predefined template, which it forwards to the assistant agent.
The assistant agent analyzes the input and selects an appropriate function call with corresponding parameters. 
The user proxy executes this call and returns the result, prompting the assistant agent to determine the next step.
This iterative exchange continues until all subtasks are completed, after which the assistant agent compiles the final summarized report. 

\noindent \textbf{Task decomposition.}
In the task\_decompose function, the user agent submits a complex task to the task decomposition agent, which decomposes it into subtasks.
These subtasks guide the subsequent processing stages.

\noindent \textbf{Retrieval and generation.}
After task decomposition, the assistant agent creates queries based on the subtasks and instructs the user proxy to call the retrieve\_generate function.
Here, the financial retrieval agent accepts a query $q$ and a number $n$, retrieves the top-$n$ transcript chunks using an embedding model for semantic similarity, and passes these chunks, along with the financial statement, to the financial manager agent.
The financial manager agent processes the inputs to answer $q$, yielding the final output.


Through this collaborative framework, the user proxy agent orchestrates the exchange of information among agents, ensuring a seamless workflow that culminates in the generation of a template-based summarized report. 
\vspace{-5pt}
\subsection{DecomposedIR}
As shown in Figure \ref{fig:decomposed}, we prompted LLM to decompose the report template into subqueries $SQ$ based on its sections using prompt chaining.
For each $SQ$, we retrieved the top-$n$ relevant chunks from the earnings call transcript using the same embedding model applied in AgenticIR. 
The embedding model encodes each chunk and the subquery into 768-dimensional embeddings. 
Relevance is determined by calculating the cosine similarity between the embedding of the subquery and the embeddings of all chunks.
These $n$ chunks, along with the financial statements, are then provided to the LLM generator. 
The LLM synthesizes the information to generate an answer for the $SQ$, combining textual context from the transcript chunks with precise numerical data from the financial statements. 
Once all $SQ$ have been processed, the answers are aggregated and summarized by section using the report template. 
The summarization process is performed by the LLM, which consolidates the answers into a coherent and comprehensive report, aligned with the structure of the original template.
\vspace{-7pt}
\section{Experiments}

For the AgenticIR and DecomposedIR, we used GPT-4o-mini \citep{gpt-4o} as the LLM.
For retrieving relevant chunks from earnings releases, we employed the \texttt{fin-mpnet-base} as the embedding model in the financial dataset. 
To ensure consistency across experiments, we uniformly set the number of retrieved chunks $n$ to 3.

To further investigate the performance across different domains, we conducted experiments on the SumIPCC \citep{SumIPCC} dataset, consisting of climate reports derived from the long-form reports of the Intergovernmental Panel on Climate Change. 
The dataset contains 140 summaries, which we reorganized into 7 reports, each comprising multiple sections and subsections.
In this dataet, we used the \texttt{stella\_en\_1.5B\_v5} as the embedding model. 

\begin{table}
\scriptsize
\centering
\caption[c]{Results on financial dataset. DE stands for DecompEval, and GE stands for G-Eval. The highest score is denoted in \textbf{bold}, and the second-highest score is \underline{underlined}.}
\vspace{-5pt}
\resizebox{\linewidth}{!}{
\begin{tabular}{lcccccccccccc}
\toprule
\multirow{2}{*}{\textbf{Methods}} & \multicolumn{2}{c}{\makecell[c]{\textbf{Financial}\\\textbf{Takeaways}}} & \multicolumn{2}{c}{\makecell[c]{\textbf{Financial}\\\textbf{Context}}} & \multicolumn{2}{c}{\makecell[c]{\textbf{Reasoning}\\\textbf{Correctness}}} & \multicolumn{2}{c}{\makecell[c]{\textbf{Management}\\\textbf{Expectation}}} \\
\cmidrule(lr){2-3} \cmidrule(lr){4-5} \cmidrule(lr){6-7} \cmidrule(lr){8-9}
 & \textbf{DE} & \textbf{GE} & \textbf{DE} & \textbf{GE} & \textbf{DE} & \textbf{GE} & \textbf{DE} & \textbf{GE} \\
\midrule
AgenticIR & 0.26 & 2.77 & 0.67 & 3.11 & 0.76 & 3.58 & 0.25 & 2.77 \\
\quad w/ sr & \underline{0.27} & 2.98 & 0.80 & 3.51 & 0.88 & 3.83 & 0.32 & 3.22 \\
DecomposedIR & 0.23 & \underline{3.47} & \underline{0.89} & \underline{3.83} & \underline{0.96} & \underline{3.97} & \underline{0.47} & \underline{3.67} \\
\quad w/ sr & \textbf{0.28} & \textbf{3.76} & \textbf{0.96} & \textbf{3.98} & \textbf{0.99} & \textbf{4.00} & \textbf{0.59} & \textbf{3.94} \\
\bottomrule
\end{tabular}
}
\vspace{-10pt}
\label{tab:DecompEval}
\end{table}

\vspace{-5pt}
\subsection{Dataset: Finance}

\noindent\textbf{Evaluation metrics.} 
Due to the predominant cost of collecting expert-written template-based financial reports, it is not feasible to conduct an apple-to-apple comparison between the generated report and a gold reference.
Therefore, we decided to adopt two LLM-based reference-free evaluation metrics, DecompEval \citep{decompeval} and G-Eval \citep{g-eval}, to assess the generated reports.
Inspired by \citet{from_fact_to_insights}, these metrics evaluate on four key characteristics: (1) Financial Takeaways: crucial financial details or numerical statistics related to the company's performance this quarter; (2) Financial Context: insights into the company's current financial performance, including references to previous quarters; (3) Reasoning Correctness: the accuracy of reasoning or explanations regarding the company's financial performance this quarter; and (4) Management Expectation: forecasts or projections about the company's future performance or expectations for the next quarter. 
In DecompEval, we prompted an LLM to evaluate whether each sentence in the generated report adheres to a specified characteristic. 
The individual sentence-level assessments are then aggregated to produce an overall score for that characteristic.
In contrast, G-Eval evaluated the generated reports by asking LLM to assign a score from 1 to 5 for each characteristic, where a score of 1 indicates poor alignment and a score of 5 represents the highest performance.

\noindent \textbf{Experimental methods.}
To further understand the performance of both frameworks, we introduced the variants employing self-reflection mechanism (denoted as w/ sr) on the four characteristics.
This mechanism allows the agent to analyze its generated content, identify the extent of alignment with each characteristics, and adjust its generated answer strategies accordingly.

\noindent \textbf{Quantitative results.}
We used GPT-4o-mini as the LLM during evaluation, and the results were shown in Table \ref{tab:DecompEval}.
We find that DecomposedIR methods consistently outperform AgenticIR methods in all metrics.
Quantitatively, DecomposedIR statistically significantly surpasses AgenticIR by 27\% on average for all characteristics with $p < 0.05$ (p-values based on Pearson's $\chi^2$ test).
This demonstrates the value of using prompt chaining to break down the report template into detailed subqueries, in line with the discovery of previous works \citep{wu2024stateflow, kwak-etal-2024-classify}.
In addition, our results show that the additional self-reflection mechanism increases DecomposedIR in DE and GE by 9.8\% on average, and AgenticIR slightly performs better as well.
This indicates the effectiveness of incorporating self-reflection into the template-based report generation task.
\begin{table}
\footnotesize
\centering
\caption[c]{Readability of both methods on financial dataset.}
\vspace{-7pt}
\begin{tabular}{lcccc}
\toprule
\textbf{Methods} & \textbf{\#Sents} & \textbf{FKGL} & \textbf{CLI} & \textbf{ARI} \\
\midrule
AgenticIR & 25.01 & 14.07 & 14.45 & 15.36 \\
\quad w/ sr & 31.82 & 15.07 & 15.80 & 16.37 \\
DecomposedIR & 77.72 & 16.52 & 16.81 & 18.07 \\
\quad w/ sr & 96.07 & 16.76 & 17.18 & 18.33 \\
\bottomrule
\end{tabular}
\vspace{-10pt}
\label{tab:readability}
\end{table}

\noindent\textbf{Readability.}
To delve deeper into the results of both frameworks, we employed three metrics: Flesch-Kincaid Grade Level (FKGL) \citep{fkgl}, the Coleman–Liau Index (CLI) \citep{cli}, and the Automated Readability Index (ARI) \citep{ari} to evaluate the readability of generated reports. 
These metrics are computed based on linguistic features such as the number of characters (CLI and ARI), syllables (FKGL), words, and sentences.
Higher scores indicate greater complexity of the reports. 
Additionally, we consider the average number of sentences per report to observe the length variation. 
In general, it can be seen that the better a framework achieves in DE and GE, the more complex the generated report becomes, as illustrated in Table \ref{tab:readability}.
Here, self-reflection mechanism not only increases the complexity of the generated report but also produces longer documents.
We hypothesize that it reflects the nature of more detailed and contextually content being more suitable for readers with higher expertise background, and statistics such as syllable counts may not be a good measure of readability for financial reports.
This aligns with prior work showing the necessity of other readability metrics \citep{financereadability} in the context of understanding companies' financial performance.
\vspace{-5pt}
\subsection{Dataset: SumIPCC}
Since SumIPCC includes ground truth data, we employed ROUGE \citep{rouge} and BERTScore \citep{bert_score} as evaluation metrics, using \texttt{deberta-v3-\\large} \cite{deberta-v3} as the BERTScore model.
As shown in Table \ref{tab:SumIPCC_Rouge}, DecomposedIR achieves the highest scores in both ROUGE and BERTScore, outperforming AgenticIR by 33\% in ROUGE-1 and 6.3\% in BERTScore. 
This suggests that DecomposedIR method enables more accurate content retrieval and summarization as well in other domains.
We conjecture that, unlike DecomposedIR, AgenticIR's autonomously derived query decomposition strategy lacks explicit structural guidance, potentially leading to inconsistencies and ultimately hindering its performance.

To investigate the effects of self-reflection in this scenarios, we further prompted LLMs with the ground-truth report of the SumIPCC dataset to come up with four characteristics similar to financial dataset.
Then, the self-reflection variants in both frameworks will conduct alignment checks with the characteristics and generate the output reports accordingly.
It can be observed that, with self-reflection, DecomposedIR outperforms AgenticIR by 45\% in ROUGE-1 and 8.3\% in BERTScore, sharing similar findings with the experiments of financial dataset.
However, we note that in this dataset, each report consists of sections and subsections with distinct focal points and stylistic variations.
Consequently, the application of self-reflection may have led to a misalignment between the generalized guide of the model and the specific requirements of each section, which eventually resulted in lower performance compared to their counterpart.

\begin{table}
\footnotesize
\caption{Results on SumIPCC dataset.}
\setlength{\tabcolsep}{3pt}
\vspace{-7pt}
\begin{tabular}{lrrrr}
\toprule
\textbf{Methods} & \textbf{ROUGE-1} & \textbf{ROUGE-2} & \textbf{ROUGE-L} & \textbf{BERTScore} \\
\midrule
AgenticIR & 0.24 & 0.05 & 0.15 & 0.63 \\
\quad w/ sr & 0.20 & 0.04 & 0.12 & 0.60 \\
DecomposedIR & \textbf{0.32} & \textbf{0.08} & \textbf{0.17} & \textbf{0.67} \\
\quad w/ sr & 0.29 & 0.06 & 0.15 & 0.65 \\
\bottomrule
\end{tabular}
\vspace{-10pt}
\label{tab:SumIPCC_Rouge}
\end{table}

\vspace{-5pt}
\section{Conclusion}
In this work, we introduce the comparison between two LLM-based workflow for template-based financial reports generation, namely AgenticIR and DecomposedIR, offering insights into the real-world framework selection.
There are two major findings in our experiments.
Firstly, decomposed prompt chaining outperforms the multi-agent approach in two evaluation scenarios, showcasing AgenticIR's limitations in bridging the gap between unstructured context and structured report template. 
Secondly, self-reflection on the key characteristics effectively improves the report quality but reduces the readability across both frameworks, highlighting its impact on report complexity.
We believe the comparison can serve as an inspiration for more research on using LLMs for template-based financial report generation, and several interesting future directions can be further investigated, such as exploring the enhancement of multi-agents with Query-Focused Summarization methods or Aspect-based Summarization in template-based report generation.



\bibliographystyle{ACM-Reference-Format}
\balance
\bibliography{reference}


\end{document}